\documentclass[11pt]{article}

\usepackage[margin=1in]{geometry}
\usepackage[T1]{fontenc}
\usepackage[utf8]{inputenc}
\usepackage{lmodern}
\usepackage{amsmath}
\usepackage{amssymb}
\usepackage{graphicx}
\graphicspath{{../figures/}}
\usepackage{booktabs}
\usepackage{natbib}
\usepackage[hidelinks]{hyperref}
\usepackage{xurl}

\title{Loop content of bond percolation on hyperbolic triangulations:\\ an exact identity and a $1/\lambda$ expansion}
\author{Zachary Treisman}
\date{\today}

\begin{document}

\maketitle

\begin{abstract}
We study the homology of percolation clusters on disk-shaped patches of the regular $\{3,q\}$ triangle tilings for $q\ge7$. Varying $q$ gives a one-parameter family of hyperbolic tilings, from mildly curved ($q=7$) to deeply hyperbolic ($q=20$), and this paper measures and derives loop content as a function of curvature. A single growth-rate parameter $\lambda(q)$, the asymptotic ratio of successive ring sizes, governs every result. Total persistent $H_1$ of the bond-percolation complex, per vertex, is measured across $q=7,\dots,20$ and found to be a nearly linear function of $1/\lambda(q)$. An exact identity reduces total loop persistence to the weight of the lattice's minimum spanning tree minus a boundary correction. The boundary correction follows from the ring recursion, and the spanning-tree weight is derived to first order in $1/\lambda(q)$ by attaching rings one at a time to a contracted interior, with a correction for loops that close through the ring outside. The resulting formula has no free parameters, gives the large-curvature intercept in closed form as $5/4-2\pi/(3\sqrt3)$, and agrees with the measurements to within the size of the next-order term.
\end{abstract}

\section{Introduction}
\label{sec:intro}

Hierarchical data is often modeled as, or embedded via, tree structure. Hyperbolic embeddings such as Poincar\'{e} or Lorentz embeddings place hierarchies in hyperbolic space specifically to preserve tree distances \citep{nickel2017}. With this context, we study the interaction between two considerations that should be taken into account when modeling real data. First, many naturally occurring branching structures are not trees. Connections can exist between downstream nodes independent of heritage. Second, restrictions on possibility and the results of random chance lead us to consider random subgraphs of an underlying graph. Percolation is the process of constructing these random subgraphs. So we study the topology of percolated subgraphs of graphs that embed in hyperbolic space, with a goal being to measure the cost of using an embedding built on the assumption that data are tree-like. Coordinates representing two paths to the same data point, embedded with a tree-like assumption, will disagree by however much the loop's holonomy fails to close.

This cost is intuitive but rarely turned into a measurable quantity for a specific graph family. This paper does that, using the regular hyperbolic tilings $\{3,q\}$ (equilateral-triangle tessellations with $q$ triangles meeting at every vertex) as a curvature dial. Varying $q$ from $7$ upward interpolates between loop-rich and more tree-like percolation behavior, through a single computable growth-rate parameter $\lambda(q)$ (\S\ref{sec:curvature-knob}). The flat triangular lattice $q=6$ is the boundary of this family, with $\lambda(6)=1$; it is excluded here for reasons given in \S\ref{sec:curvature-knob}.

Section~\ref{sec:partA} treats the topology of the percolated complex. Bond percolation defines a filtered simplicial complex, and its total persistent $H_1$ (summed bar length across the whole percolation process, per vertex) is measured across eleven values of $q$. The main result is a derivation of that quantity. An exact identity, closely related to known relations between persistence lifetimes and minimum spanning acycles \citep{hiraoka2017,skraba2020}, reduces total $H_1$ persistence to the weight of the lattice's minimum spanning tree minus a boundary correction. The boundary correction is a consequence of the ring recursion that defines $\lambda(q)$. The spanning-tree weight is derived by attaching rings one at a time to a contracted interior, which reduces it to a percolation calculation on a cycle with spokes, plus a correction for the rare but systematic loops that close through the ring outside. The formula that results has no adjustable parameters and matches the measurements to within the size of the next-order term. In particular it explains why the measured values are so close to linear in $1/\lambda(q)$: the expansion parameter of the derivation is $1/\lambda(q)$ itself.

Section~\ref{sec:diagnostic} sets up persistent homology as the diagnostic used throughout. Section~\ref{sec:curvature-knob} defines the family, its growth rate $\lambda(q)$, and the combinatorial structure of its rings that the derivation depends on. Section~\ref{sec:partA} contains the derivation and the measurements, and Section~\ref{sec:conclusion} summarizes and lists open questions.

\section{Persistent homology as the diagnostic}
\label{sec:diagnostic}

Bond percolation defines a natural filtration. Given a graph $G$, assign each edge $e\in G$ an i.i.d.\ Uniform$[0,1]$ occupation threshold $u_e\in[0,1]$, and build the clique (flag) complex of the graph using those thresholds as filtration values. Declare edge $e$ occupied at parameter $p$ whenever its threshold $u_e\le p$, and declare 2-simplices occupied wherever a triangle's three edges are all present. A triangle enters the filtration at the maximum of its three edge thresholds. Reusing the same draw of thresholds across all $p$ makes the occupied sets nested: one draw of $\{u_e\}$ gives a single filtration that encodes the entire coupled percolation process across every value of $p$. Standard persistent homology software (here, GUDHI) extracts persistent $H_0$ and $H_1$ from that one filtration directly, as Betti curves $\beta_0(p)$, $\beta_1(p)$ and persistence diagrams whose bars record when each loop is born and when it gets filled in. Any single draw of $\{u_e\}$ is still one sample path, and the resulting curves fluctuate from draw to draw; the measurements below average over 15 independent draws per lattice. This connects to existing work on homological percolation, which studies the birth and death of giant cycles in percolation processes via the Euler characteristic curve on cubical and other lattices \citep{bobrowski2020}; the approach here specializes to the $\{3,q\}$ triangulated case and adds curvature as an explicit parameter.

One technical condition makes this tractable: as long as the graph's only 3-cliques are genuine faces of the underlying triangulation (no ``accidental'' triangles among mutually adjacent vertices that don't bound a real 2-cell), the clique complex reconstructs the correct 2-cell structure automatically. This was verified computationally for every lattice used below (Euler characteristic exactly 1 in each case, confirming a topological disk, and an exact match between the graph's 3-cliques and the known face list, with zero spurious or missing triangles).

\begin{figure}[htbp]
\centering
\begin{tabular}{cc}
\includegraphics[width=0.48\textwidth]{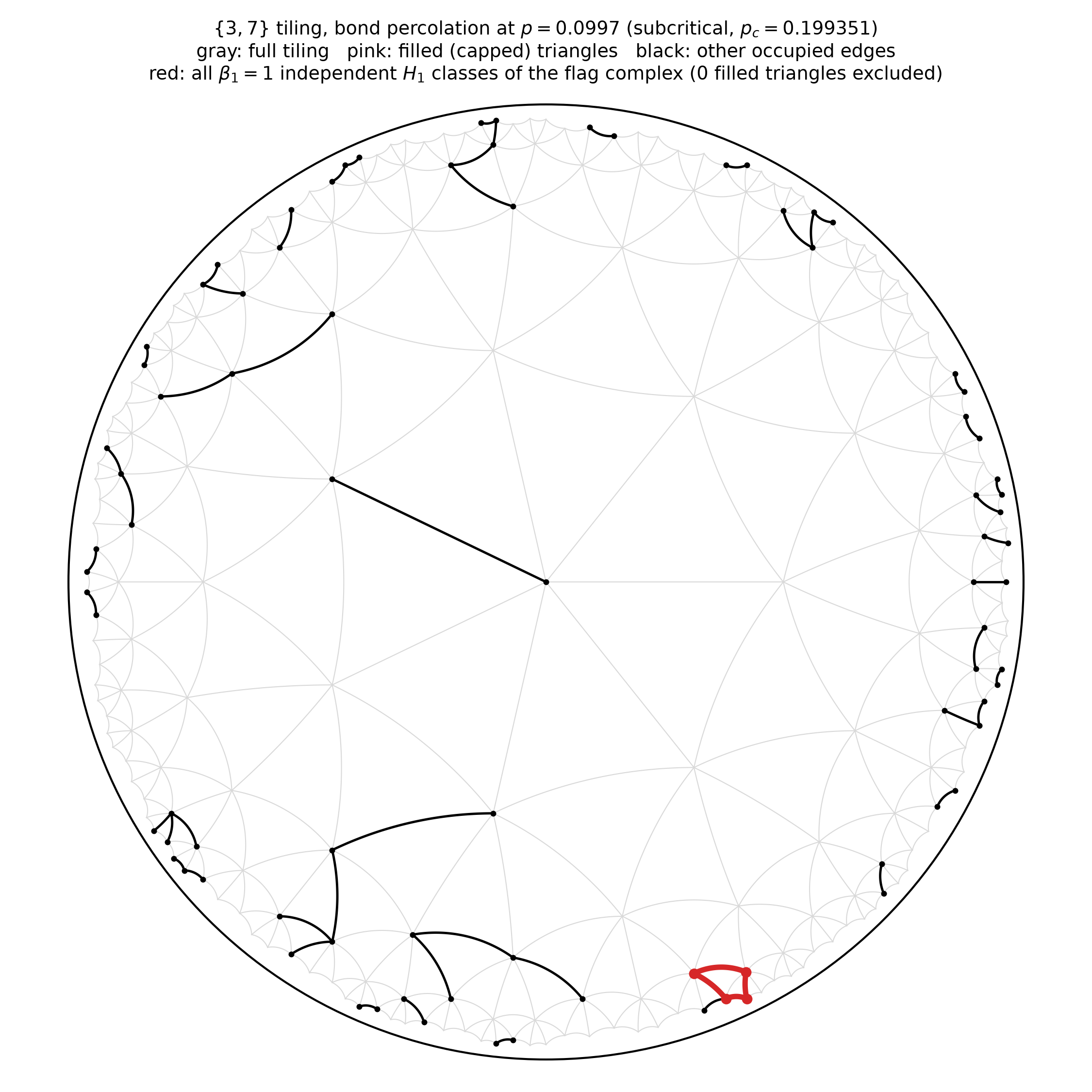} &
\includegraphics[width=0.48\textwidth]{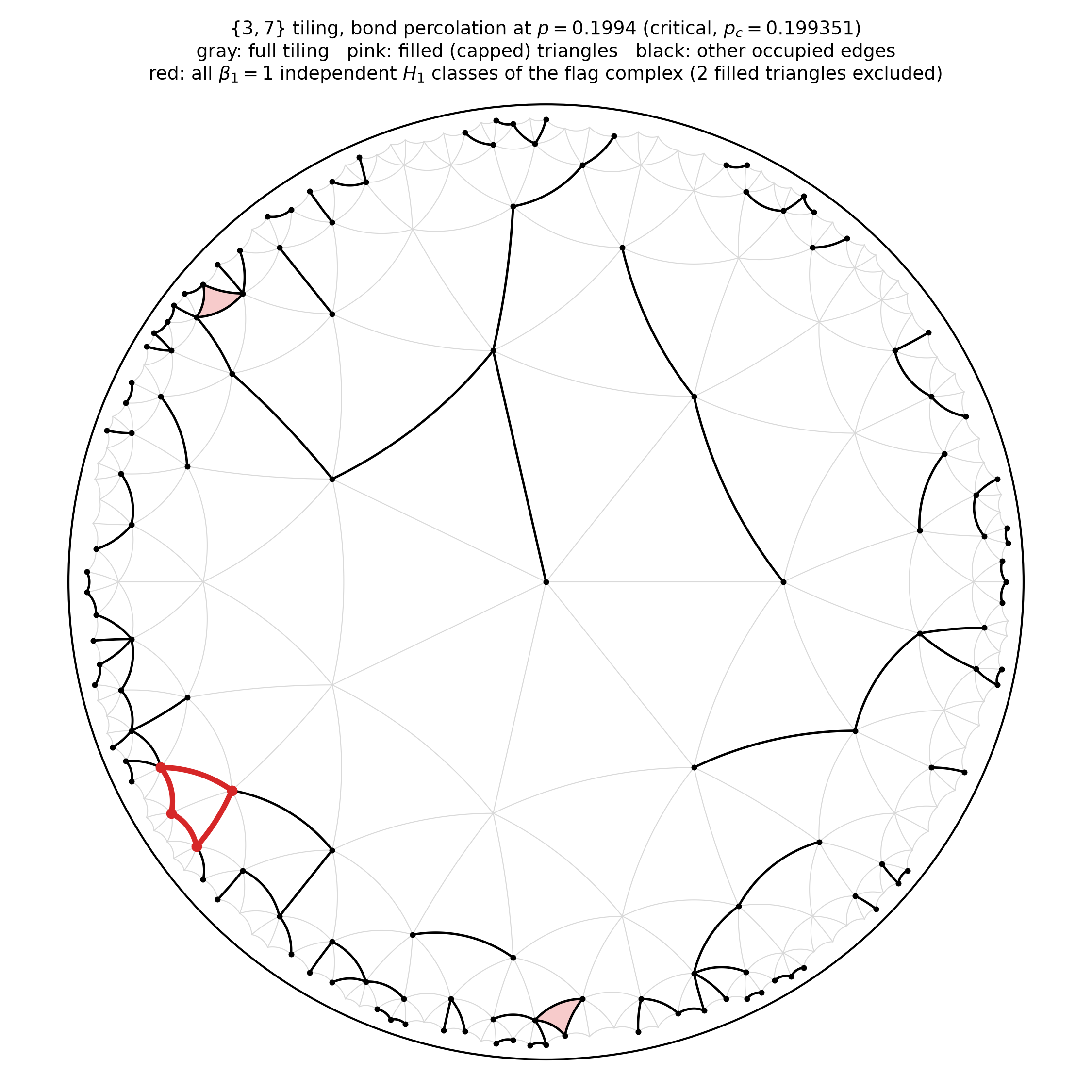} \\
\includegraphics[width=0.48\textwidth]{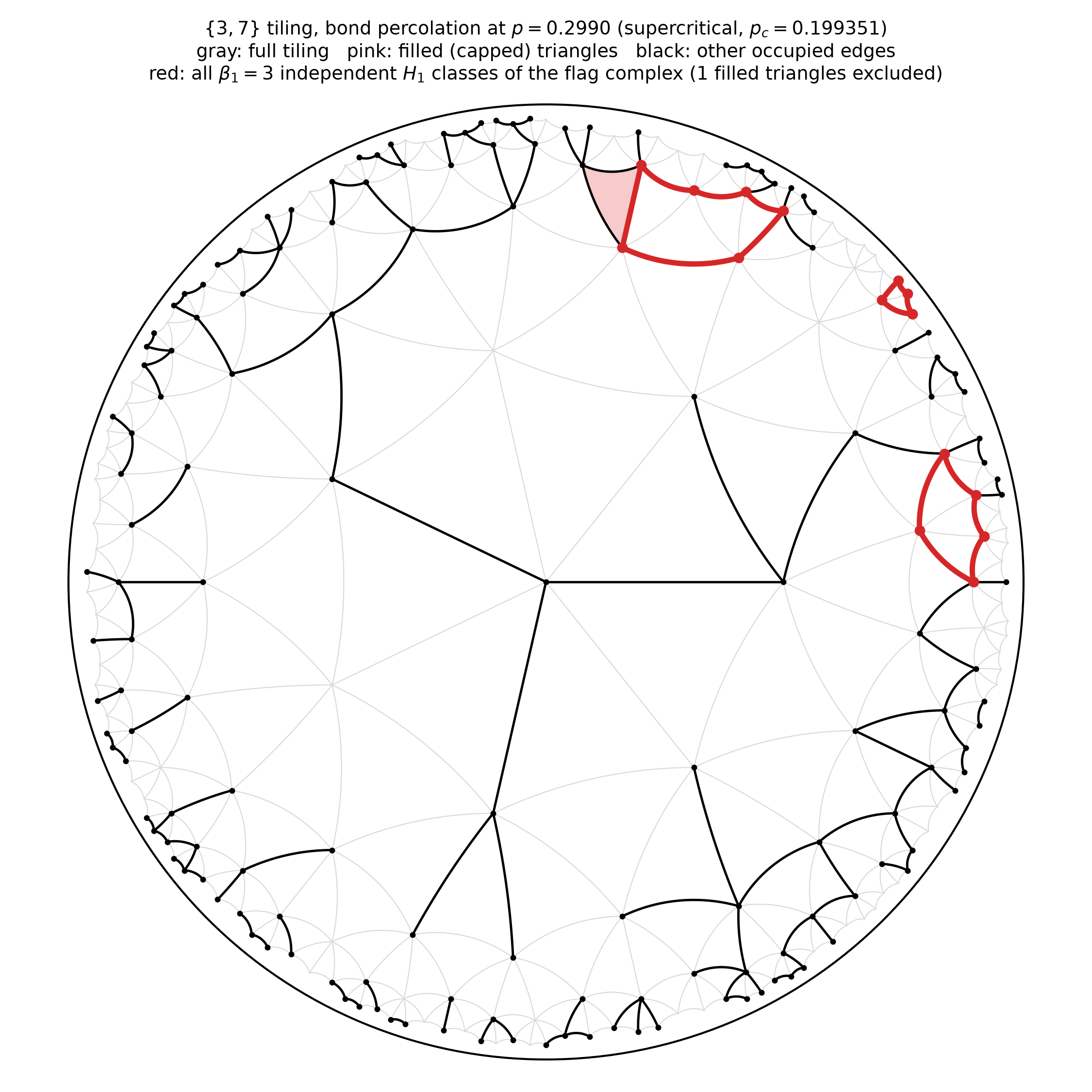} &
\includegraphics[width=0.48\textwidth]{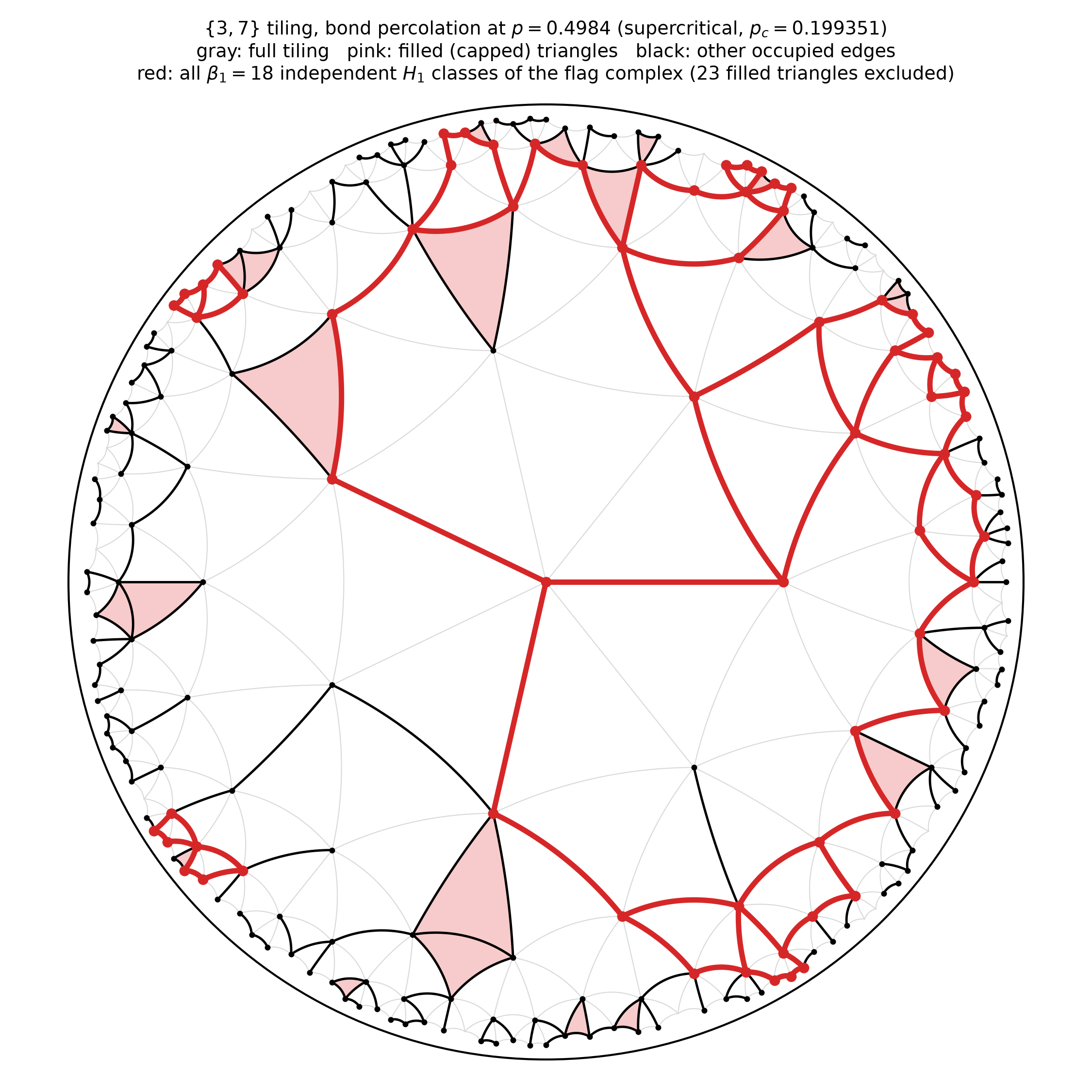}
\end{tabular}
\caption{The $\{3,7\}$ tiling to ring 4 ($V=232$) in the Poincar\'{e} disk, with every edge drawn as a hyperbolic geodesic, under bond percolation at $p=0.10$, $0.20$, $0.30$, and $0.50$ (top left to bottom right; the bond threshold is $p_c\approx0.199$). Gray: the full tiling. Black: occupied edges. Pink: triangles with all three edges occupied, which the flag complex fills in. Red: a basis for $H_1$ of the flag complex, computed as the cycle space of the occupied graph modulo the boundaries of the filled triangles. At this small size the critical draw happens to carry a single loop; loop density per vertex at $p_c$ is small but nonzero, and grows with $p$ well past the threshold, which is the regime that dominates the integrated quantity studied in \S\ref{sec:partA}.}
\label{fig:poincare}
\end{figure}

One subtlety matters again in \S\ref{sec:identity}. A single triangular face's own 3-cycle is born, in this filtration, only once all three of its edges are present, which is also the threshold at which the 2-simplex fills it in. Isolated triangle loops are therefore born and die simultaneously and contribute zero persistence; whatever loop content survives has to come from cycles spanning more than one face. Figure~\ref{fig:poincare} shows what this looks like on $\{3,7\}$ at four occupation probabilities: the filled triangles are shaded, and the highlighted cycles are a basis for $H_1$ of the flag complex at that $p$, so a fully occupied triangle on its own is never highlighted.

\section{The \texorpdfstring{$\{3,q\}$}{\{3,q\}} tilings and their ring structure}
\label{sec:curvature-knob}

The family of regular triangulations $\{3,q\}$ ($q$ equilateral triangles meeting at every vertex) is built by a ring-growth construction: starting from a single vertex, each successive ring is attached according to the fixed combinatorial rule that $q$ triangles meet at every vertex. Ring $n$ has $r_n$ vertices, with $r_1=q$, $r_2=q(q-4)$, and
\[
r_n = (q-4)\,r_{n-1} - r_{n-2}, \qquad n\ge3.
\]
The dominant root of the characteristic equation is
\[
\lambda(q) = \frac{(q-4) + \sqrt{(q-4)^2-4}}{2},
\]
the tiling's asymptotic per-ring growth rate: $r_n/r_{n-1}\to\lambda(q)$ as $n\to\infty$. For $q\ge7$ the discriminant is positive and $\lambda>1$, giving exponential growth. Table~\ref{tab:ladder} gives illustrative values.

\begin{table}[htbp]
\centering
\begin{tabular}{llr}
\toprule
$q$ & geometry & $\lambda(q)$ \\
\midrule
7  & mildly hyperbolic & 2.618 \\
8  & more hyperbolic & 3.732 \\
12 & strongly hyperbolic & 8.873 \\
20 & deep hyperbolic & 17.944 \\
\bottomrule
\end{tabular}
\caption{Illustrative values of the growth rate $\lambda(q)$.}
\label{tab:ladder}
\end{table}

Two consequences of the recursion do all the work in \S\ref{sec:partA}.

\paragraph{The outermost ring is a fixed fraction of the disk.} The total vertex count $V=1+\sum_n r_n$ is a geometric series dominated by its last term, so the number $b$ of boundary edges (edges bounding only one face) satisfies $b/V\to1-1/\lambda(q)$ as the ring depth grows. Table~\ref{tab:boundary} shows this is already accurate to four decimal places at $V\sim10^3$. A disk of any depth is therefore mostly rim: for $q=7$, $62\%$ of all vertices lie on the outermost ring, and for $q=20$, $94\%$.

\begin{table}[htbp]
\centering
\begin{tabular}{rrrr}
\toprule
$q$ & $V$ & $b/V$ & $1-1/\lambda(q)$ \\
\midrule
7  & 617     & 0.6240 & 0.6180 \\
8  & 2{,}281 & 0.7330 & 0.7321 \\
9  & 1{,}306 & 0.7925 & 0.7913 \\
12 & 6{,}817 & 0.8731 & 0.8730 \\
16 & 2{,}497 & 0.9163 & 0.9161 \\
20 & 5{,}441 & 0.9373 & 0.9373 \\
\bottomrule
\end{tabular}
\caption{Boundary-edge fraction versus its large-$V$ limit, at moderate ring depths.}
\label{tab:boundary}
\end{table}

\paragraph{Each ring has two kinds of vertex.} A vertex in ring $n$ is adjacent either to one vertex of ring $n-1$ (call it e-type) or to two (v-type). The v-type vertices are the children shared between adjacent parents, and counting them is immediate: ring $n-1$ is a cycle, so it has $r_{n-1}$ adjacent pairs, each pair has exactly one common child in ring $n$, and distinct pairs have distinct common children. Hence
\begin{equation}
v_n = r_{n-1},
\qquad\text{so}\qquad
\frac{v_n}{r_n} = \frac{r_{n-1}}{r_n} \;\longrightarrow\; \frac{1}{\lambda(q)} .
\label{eq:vtype}
\end{equation}
The arrangement of the two types around a ring is equally simple. Reading each ring as a cyclic word in the letters $e$ and $V$, ring $n$ is obtained from ring $n-1$ by the substitution
\begin{equation}
\sigma:\quad e\;\mapsto\; V\,e^{\,q-5},\qquad V\;\mapsto\; V\,e^{\,q-6},
\label{eq:substitution}
\end{equation}
so that ring $n$ is $\sigma^{\,n-1}(e^q)$, since ring 1 consists of $q$ e-type vertices. The block assigned to each parent is the child it shares with its predecessor on the ring, followed by its unshared children. An e-type vertex in ring $n-1$ has one edge inward and two along the ring, so $q-3$ edges outward, two of them to shared children; that leaves $q-5$ unshared children, and $q-6$ for a v-type vertex, which has one more edge inward. Counting letters in \eqref{eq:substitution} recovers both the ring recursion and \eqref{eq:vtype}. Because the seed $e^q$ is $q$ copies of one letter, every ring is $q$ copies of a single block, which is the $q$-fold rotational symmetry of the construction. The substitution was checked against the constructed graphs for every $q$ and ring depth used here.

\paragraph{Why $q=6$ is excluded.} At $q=6$ the recursion has a repeated root $\lambda=1$: ring sizes grow linearly, the boundary fraction $b/V$ tends to zero rather than to a positive constant, and there is no small parameter. Every result in \S\ref{sec:partA} is an expansion in $1/\lambda(q)$, and at $\lambda=1$ that expansion has nothing to expand in. The flat case also behaves differently in a purely practical sense: its per-vertex loop content converges to its large-$V$ limit only as a power law in $V$ (roughly $V^{-0.45}$, needing on the order of $10^6$ vertices for three-figure accuracy), whereas every hyperbolic $q$ is converged to within 1 to 2\% by $10^4$ vertices, consistent with the outermost ring being a fixed fraction of the whole so that adding a ring adds the same per-vertex contribution. The flat lattice is the amenable member of the family and is genuinely a different object; it is not part of this paper.

\begin{figure}[htbp]
\centering
\includegraphics[width=0.78\textwidth]{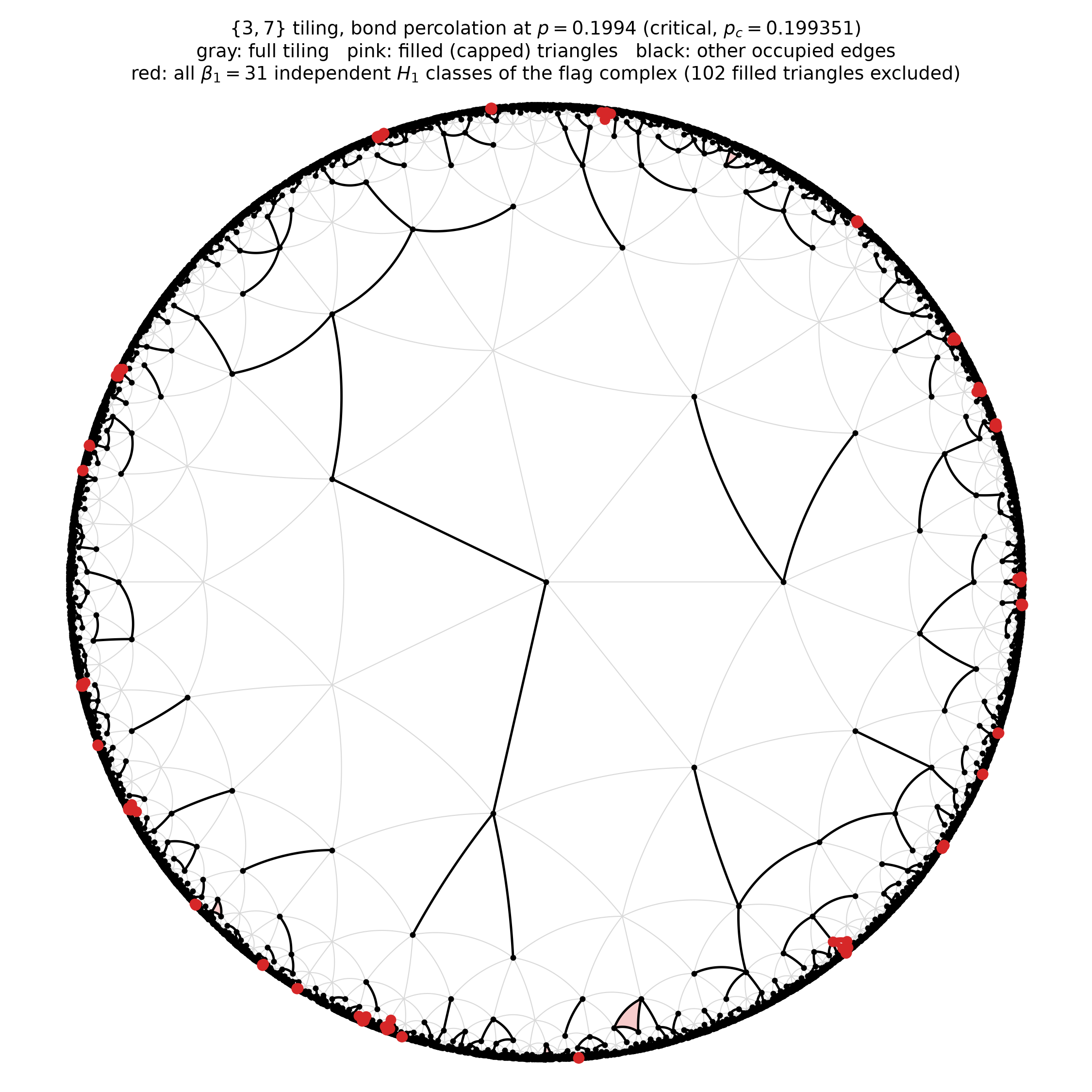}
\caption{The $\{3,7\}$ tiling to ring 8 ($V=11{,}173$) at $p=p_c$, same color scheme as Figure~\ref{fig:poincare}. This draw carries 31 independent $H_1$ classes; the loop density at $p_c$ is about $0.003$ per vertex and is stable from ring 7 onward, so the single loop in the ring-4 panel of Figure~\ref{fig:poincare} is a small-sample effect. The classes sit almost entirely on the outermost ring, which holds $62\%$ of all vertices at $q=7$. The interior looks empty because it holds few vertices, not because its loop density is lower.}
\label{fig:ring8}
\end{figure}

\paragraph{Criticality is not what is being measured.} For $q\ge7$ the tilings are nonamenable and Gromov-hyperbolic, and percolation on them has mean-field critical behavior \citep{hutchcroft2019,mertens2017}. That is not what governs the quantity studied here. Total persistent $H_1$ integrates $\beta_1(p)$ over all $p\in[0,1]$, and almost none of that integral comes from near the percolation threshold: restricting to a window $[p_c/2,\,2p_c]$ around the bond threshold captures under $5\%$ of the total at $q=7$ and under $0.01\%$ by $q=20$. Loop content in these complexes is a supercritical, bulk phenomenon, and the derivation in \S\ref{sec:partA} never uses critical-point theory. Figure~\ref{fig:ring8} shows a critical draw on a larger disk: loops are present at a small, stable density, and they sit where the vertices are, on the rim.

\section{The topological loop content of percolated subsets}
\label{sec:partA}

\subsection{An exact identity}
\label{sec:identity}

Because the percolation complex is a triangulated topological disk (\S\ref{sec:diagnostic}), $\beta_2(p)=0$ identically. Consequently the Euler characteristic $\chi(p)=V-E(p)+F(p)$ (all vertices are present at every $p$; $E(p),F(p)$ count edges and triangles occupied by threshold $p$) satisfies $\chi(p)=\beta_0(p)-\beta_1(p)$ exactly, at every $p$, giving
\[
\beta_1(p) = \beta_0(p) - V + E(p) - F(p).
\]
Total $H_1$ persistence is $\int_0^1\beta_1(p)\,dp$, the area under the loop-density curve. Integrating term by term:
\begin{itemize}
\item $u_e$ is edge $e$'s occupation threshold, $E(p)=|\{e\,|\,u_e\le p\}|$ gives $\int_0^1E(p)\,dp=\sum_e(1-u_e)$;
\item writing $m_t=\max(u_e\,|\, e\in t)$ for a triangle $t$'s occupation threshold, $\int_0^1F(p)\,dp=\sum_t(1-m_t)$;
\item for the $\beta_0(p)$ term, let $T$ be the minimum spanning tree (MST) of $G$ under the weights $u_e$, the set of $V-1$ edges connecting all vertices at minimum total weight, and let $G_p$ and $T_p$ be the subgraphs of edges with weight at most $p$. Then $G_p$ and $T_p$ have the same connected components. Since $T_p\subseteq G_p$, every component of $T_p$ lies in one of $G_p$. Conversely, if an edge $e=xy$ of $G_p$ is not in $T$, the cycle property of minimum spanning trees says every edge on the path from $x$ to $y$ in $T$ is lighter than $u_e\le p$, so $x$ and $y$ are already joined in $T_p$. Hence $V-\beta_0(p)$ is the number of MST edges of weight at most $p$, and $\int_0^1\beta_0(p)\,dp = 1+w_{\mathrm{MST}}$, where $w_{\mathrm{MST}}=\sum_{e\in T}u_e$.
\end{itemize}
Hence, for every realization,
\begin{equation}
\int_0^1\beta_1(p)\,dp = w_{\mathrm{MST}} - \sum_e u_e + \sum_t m_t.
\label{eq:exact-identity}
\end{equation}
Identities of this kind, expressing total persistence through a minimum spanning tree or its higher-dimensional analogue, the minimum spanning acycle, are known for random simplicial complexes \citep{hiraoka2017,skraba2020}; \eqref{eq:exact-identity} is the version for a filtered triangulated disk.

\subsection{Reduction to a spanning-tree weight and a boundary term}
\label{sec:boundary}

Taking expectations, $\mathbb{E}[u_e]=\tfrac12$ and $\mathbb{E}[m_t]=\tfrac34$ (the mean of the maximum of three i.i.d.\ Uniform$[0,1]$ variables), so $\mathbb{E}\big[\sum_e u_e\big]=E_{\mathrm{tot}}/2$ and $\mathbb{E}\big[\sum_t m_t\big]=3F_{\mathrm{tot}}/4$, where $E_{\mathrm{tot}},F_{\mathrm{tot}}$ are the lattice's total edge and face counts. Writing $b$ for the number of boundary edges, Euler's formula together with the disk's face structure gives the exact relations $E_{\mathrm{tot}}=3V-3-b$ and $F_{\mathrm{tot}}=2V-2-b$. Substituting into \eqref{eq:exact-identity} in expectation:
\begin{equation}
\mathbb{E}\left[\int_0^1\beta_1(p)\,dp\right] = \mathbb{E}[w_{\mathrm{MST}}] - \frac{b}{4}.
\label{eq:reduced-identity}
\end{equation}
With $b/V\to1-1/\lambda(q)$ from \S\ref{sec:curvature-knob}, the boundary term contributes exactly $-\tfrac14+\tfrac{1}{4\lambda(q)}$ to $H_1$ persistence per vertex. Everything else is in the bulk term
\[
\mu(q) := \lim_{V\to\infty}\frac{\mathbb{E}[w_{\mathrm{MST}}]}{V},
\]
the expected weight of a minimum spanning tree per vertex under i.i.d.\ Uniform$[0,1]$ edge weights. Because the outermost ring is a fixed fraction of the disk, $\mu(q)$ is a property of the disk sequence, not of a translation-invariant bulk; that is what makes it derivable.

\subsection{Deriving the bulk term}
\label{sec:bulk}

The identity $\int_0^1\beta_0(p)\,dp=1+w_{\mathrm{MST}}$ runs both ways, so $\mathbb{E}[w_{\mathrm{MST}}]=\int_0^1(\mathbb{E}[C(p)]-1)\,dp$ where $C(p)$ is the number of connected components at threshold $p$. This turns the spanning-tree weight into a percolation question: how many components, on average, at each $p$.

\paragraph{One ring at a time.} Build the disk ring by ring and ask how many new components ring $n$ contributes at threshold $p$, given everything inside it. If the interior were a single connected component, then ring $n$'s contribution would be the number of maximal runs of consecutive ring-$n$ vertices, joined by present rim edges, that have no present spoke into the interior. Ring $n$ is a cycle of $r_n$ vertices; each vertex has one spoke inward if e-type and two if v-type; rim edges and spokes are independent. The vertex types are not independent, since their arrangement is fixed by \eqref{eq:substitution}. But the calculation is an expansion to first order in $\alpha=1/\lambda(q)$, and to that order a run of $L$ vertices only sees the expected number of v-types it contains, which is $L\alpha$ whatever the arrangement. The arrangement matters only for runs containing two or more v-types, which is an $O(\alpha^2)$ effect, the same order as the terms already neglected below. To first order the types may therefore be treated as independent with probability $\alpha$. The probability that a given vertex has no present inward spoke is then $(1-p)(1-\alpha p)$, and the runs are a renewal process along the cycle with geometric lengths. Summing over runs gives, per vertex of the ring,
\begin{equation}
c(p;\alpha) = \frac{(1-p)^3\,(1-\alpha p)}{1-p(1-p)(1-\alpha p)},
\qquad
\mu_{\mathrm{wheel}}(\alpha) := \int_0^1 c(p;\alpha)\,dp.
\label{eq:wheel}
\end{equation}
Since every ring contributes the same per-vertex amount, $\mu_{\mathrm{wheel}}(\alpha)$ is the bulk density under the single-connected-interior assumption. We keep $\mu_{\mathrm{wheel}}$ as a function of $\alpha$ rather than linearizing it, which costs nothing and is more accurate at small $q$; replacing the independent types by the true arrangement changes it by less than $10^{-3}$ at $q=7$. At $\alpha\to0$ the integral is elementary:
\begin{equation}
\mu_{\mathrm{wheel}}(0) = \frac32 - \frac{2\pi}{3\sqrt3} = 0.290800\ldots
\label{eq:intercept}
\end{equation}

\paragraph{Where the assumption is exact, and where it is not.} The outermost ring has nothing outside it, so the only way one of its spoke-free runs can join anything is through the interior, and for it the assumption is exact. An inner ring is different: two of its spoke-free runs, separated by an absent rim edge between vertices $u$ and $u+1$, can be joined through the ring outside, because $u$ and $u+1$ have a common child $w$ there (the v-type vertex they share). If a run of present rim edges through $w$ carries a present spoke to $u$ on one side and to $u+1$ on the other, the two runs are one component, and the single-interior count is one too high. Averaging over the geometric length of the run through $w$ gives the probability that a given absent rim edge is bridged this way,
\begin{equation}
\beta(p) = \left(\frac{p}{1-p+p^2}\right)^2,
\label{eq:beta}
\end{equation}
so that the effective probability two consecutive ring vertices are connected, directly or via the ring outside, is $p_{\mathrm{eff}}=p+(1-p)\beta(p)$. Recomputing the run count with $p_{\mathrm{eff}}$ for rim connectivity and $p$ for spoke presence,
\[
c_{\mathrm{eff}}(p) = \frac{(1-p_{\mathrm{eff}})^2(1-p)}{1-p_{\mathrm{eff}}(1-p)},
\]
and the overcount per inner-ring vertex, integrated over $p$, is
\begin{equation}
B_1 = \int_0^1\big[c(p;0)-c_{\mathrm{eff}}(p)\big]\,dp = 0.04713\ldots
\label{eq:B1}
\end{equation}
The integrand is the rational function $p^2(1-p)^3(p^4-p^3+p^2-p+1)/[(p^2-p+1)^2(p^6-3p^5+5p^4-5p^3+5p^2-3p+1)]$; the sextic is palindromic and reduces to an irreducible cubic in $p+1/p$, so $B_1$ has a closed form, but not an illuminating one. Since the inner rings are a $1/\lambda$ fraction of the disk, and bridging through two or more rings is suppressed by further powers of $1/\lambda$,
\begin{equation}
\mu(q) = \mu_{\mathrm{wheel}}\!\left(\tfrac{1}{\lambda(q)}\right) - \frac{B_1}{\lambda(q)} + O\!\left(\lambda(q)^{-2}\right).
\label{eq:mu-derived}
\end{equation}

\paragraph{Checks on the pieces.} Each ingredient was tested separately by simulation on the constructed graphs, not only the final sum. The outermost ring's run count matches $c(p;\alpha)$ to about $1\%$ at every $p$, and its overcount is exactly zero at every $p$. The bridging probability $\beta(p)$, measured directly at absent rim edges, matches \eqref{eq:beta} to within $2\%$ across $p$. The per-vertex overcount of an inner ring, measured at $q=20$ and $q=12$ where the ring outside is long enough for the local picture to apply, integrates to $0.047$ and $0.045$ against $B_1=0.0471$. The abstract run-counting model reproduces $c-c_{\mathrm{eff}}$ to four digits in its own Monte Carlo.

\subsection{The predicted law and the measurements}
\label{sec:measurements}

Combining \eqref{eq:reduced-identity} and \eqref{eq:mu-derived}, total persistent $H_1$ per vertex is
\begin{equation}
\frac{H_1}{V} = \mu_{\mathrm{wheel}}\!\left(\tfrac1\lambda\right) - \frac{B_1}{\lambda} - \frac14 + \frac{1}{4\lambda} + O(\lambda^{-2}),
\qquad
\lim_{\lambda\to\infty}\frac{H_1}{V} = \frac54 - \frac{2\pi}{3\sqrt3} = 0.04080\ldots
\label{eq:h1-derived}
\end{equation}
There are no fitted parameters in \eqref{eq:h1-derived}. The large-curvature intercept is a closed form. The dependence on $q$ enters only through $\lambda(q)$, and the leading behavior is linear in $1/\lambda$ because that is the expansion parameter; $\mu_{\mathrm{wheel}}$ is not exactly linear in $\alpha$, which is why a straight line is a good but not perfect description.

\begin{figure}[htbp]
\centering
\includegraphics[width=\textwidth]{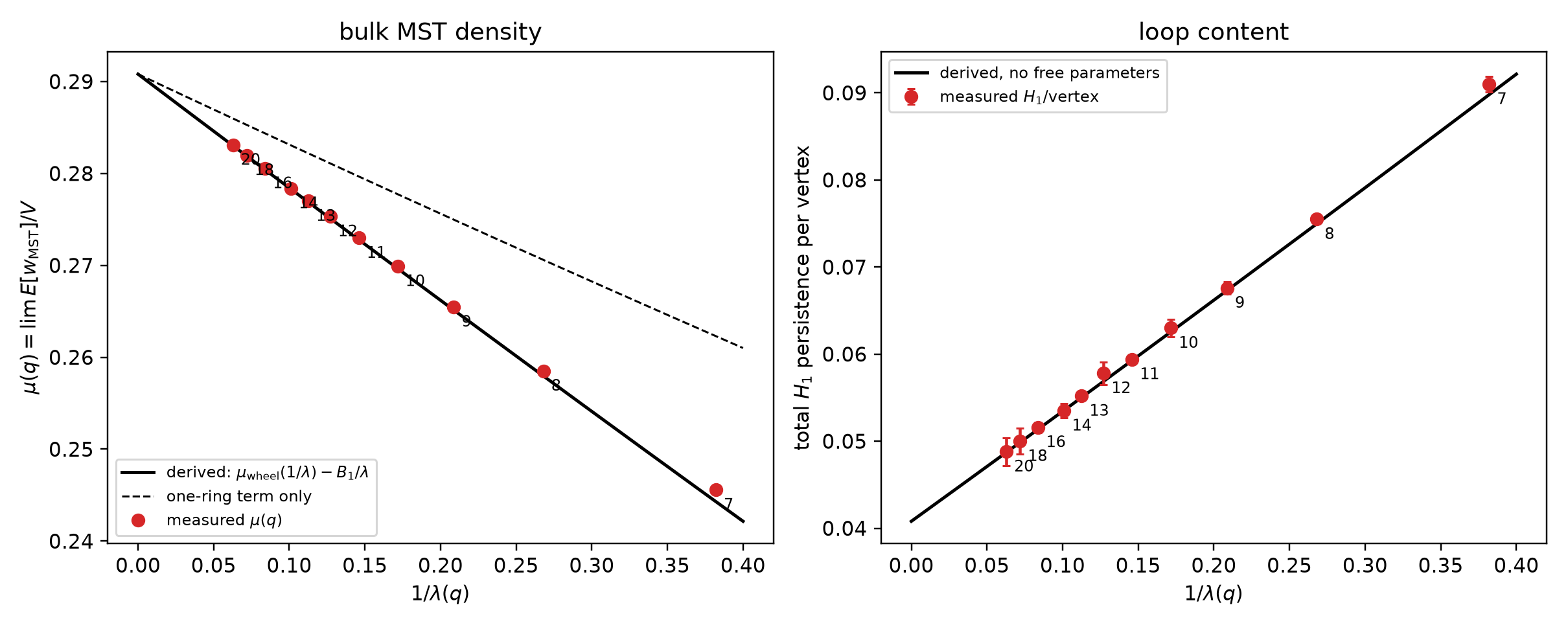}
\caption{Left: measured $\mu(q)$ against $1/\lambda(q)$, with the derived curve \eqref{eq:mu-derived} (solid) and the one-ring term $\mu_{\mathrm{wheel}}$ alone (dashed). Right: measured total $H_1$ per vertex with the derived curve \eqref{eq:h1-derived}. Neither curve has a fitted parameter.}
\label{fig:derived}
\end{figure}

\paragraph{Measurements.} Total persistent $H_1$ was measured for $q=7,8,9,10,11,12,13,14,16,18,20$, at each $q$'s largest tested system size ($V$ between $3{,}781$ and $31{,}809$), averaging 15 percolation draws. The bulk density $\mu(q)$ needs only Kruskal's algorithm, so it was measured separately at much larger sizes, up to $V\approx2\times10^6$, averaging 20 draws. Tables~\ref{tab:mu} and~\ref{tab:h1} compare both to the derived values, and Figure~\ref{fig:derived} plots them against $1/\lambda(q)$.

\begin{table}[htbp]
\centering
\begin{tabular}{rrrrrr}
\toprule
$q$ & $\lambda(q)$ & measured $\mu(q)$ & derived, eq.~\eqref{eq:mu-derived} & difference & difference$\,\times\lambda^2$ \\
\midrule
7 & 2.618 & 0.24557 & 0.24430 & $-0.00127$ & $-0.009$ \\
8 & 3.732 & 0.25852 & 0.25796 & $-0.00056$ & $-0.008$ \\
9 & 4.791 & 0.26548 & 0.26514 & $-0.00034$ & $-0.008$ \\
10 & 5.828 & 0.26990 & 0.26966 & $-0.00024$ & $-0.008$ \\
11 & 6.854 & 0.27297 & 0.27279 & $-0.00018$ & $-0.008$ \\
12 & 7.873 & 0.27532 & 0.27511 & $-0.00021$ & $-0.013$ \\
13 & 8.887 & 0.27700 & 0.27688 & $-0.00012$ & $-0.009$ \\
14 & 9.899 & 0.27837 & 0.27830 & $-0.00007$ & $-0.007$ \\
16 & 11.916 & 0.28055 & 0.28040 & $-0.00015$ & $-0.021$ \\
18 & 13.928 & 0.28198 & 0.28190 & $-0.00008$ & $-0.016$ \\
20 & 15.937 & 0.28310 & 0.28302 & $-0.00008$ & $-0.020$ \\
\bottomrule
\end{tabular}
\caption{Bulk minimum-spanning-tree weight per vertex, measured at $V$ up to $2\times10^6$, against the derived value. Standard errors on the measured values are $4$ to $7\times10^{-5}$. For $q\le11$ the differences are many standard errors and scale as $\lambda^{-2}$ with coefficient near $-0.01$, the neglected term in \eqref{eq:mu-derived}; for larger $q$ they are within two or three standard errors of that.}
\label{tab:mu}
\end{table}

\begin{table}[htbp]
\centering
\begin{tabular}{rrrrr}
\toprule
$q$ & $V$ & measured $H_1/V$ & derived, eq.~\eqref{eq:h1-derived} & difference \\
\midrule
7 & 29{,}261 & $0.0910\pm0.0002$ & 0.0898 & $-0.0012$ \\
8 & 31{,}809 & $0.0755\pm0.0001$ & 0.0750 & $-0.0005$ \\
9 & 30{,}025 & $0.0676\pm0.0002$ & 0.0673 & $-0.0003$ \\
10 & 14{,}351 & $0.0630\pm0.0003$ & 0.0625 & $-0.0005$ \\
11 & 29{,}041 & $0.0594\pm0.0001$ & 0.0593 & $-0.0001$ \\
12 & 6{,}817 & $0.0578\pm0.0003$ & 0.0569 & $-0.0009$ \\
13 & 10{,}414 & $0.0552\pm0.0001$ & 0.0550 & $-0.0002$ \\
14 & 15{,}261 & $0.0535\pm0.0002$ & 0.0536 & $+0.0001$ \\
16 & 29{,}761 & $0.0516\pm0.0002$ & 0.0514 & $-0.0002$ \\
18 & 3{,}781 & $0.0500\pm0.0004$ & 0.0498 & $-0.0002$ \\
20 & 5{,}441 & $0.0488\pm0.0004$ & 0.0487 & $-0.0001$ \\
\bottomrule
\end{tabular}
\caption{Total persistent $H_1$ per vertex, measured (mean and standard error over 15 draws) against the derived value. Root-mean-square difference $0.0005$.}
\label{tab:h1}
\end{table}

Since the boundary term is exact, the $H_1$ differences should track the $\mu$ differences, and they do: $-0.0012$ against $-0.0013$ at $q=7$, $-0.0005$ against $-0.0006$ at $q=8$, and within a few $10^{-4}$ elsewhere. At $q=7$ and $8$ the difference is many standard errors, and it is the $O(\lambda^{-2})$ term. The largest remaining discrepancy, at $q=12$, is about two standard errors from the $\mu$ difference, at one of the smallest system sizes measured.

\paragraph{On fitting.} Before the derivation existed, the natural summary of Table~\ref{tab:h1} was a two-parameter fit, $H_1/V\approx0.0404+0.1316/\lambda(q)$ with $R^2=0.9995$. A fit that good over eleven points of a smooth monotone function is weaker evidence than it feels: many decreasing functions of $q$ would fit comparably, and $1/\lambda(q)$ could have been an artifact of the range. The derivation removes that concern by producing the functional form rather than assuming it. A straight line fitted to the derived curve \eqref{eq:h1-derived} over the same $q$ is $0.0406+0.1286/\lambda(q)$, close to the empirical fit. The fit is now a check on the derivation, not the result.

\subsection{Remarks}
\label{sec:remarks}

The large-curvature limit of the bulk term, $\mu_{\mathrm{wheel}}(0)=3/2-2\pi/(3\sqrt3)$, should not be read as a treelike or Bethe-lattice value. As $q\to\infty$ the v-type fraction $1/\lambda$ goes to zero, so each ring attaches to the interior through single spokes, but the ring is still a cycle of filled triangles, and the cycle is what the calculation in \S\ref{sec:bulk} is about. The tiling never becomes the triangle-free regular tree that ``Bethe lattice'' ordinarily denotes; it is tree-like at long range only, and the nonzero intercept $5/4-2\pi/(3\sqrt3)$ in \eqref{eq:h1-derived} is the loop content that survives on the rim.

The expected weight of a random minimum spanning tree is a classical object. On the complete graph it converges to $\zeta(3)$ \citep{frieze1985}, a result most cleanly proved through the local weak limit, the Poisson-weighted infinite tree \citep{aldous2004}; on infinite nonamenable graphs the relevant objects are the free and wired minimal spanning forests \citep{lyons2006}. The disks here are neither: their outermost ring is a fixed fraction of the whole, which is exactly what lets a ring-by-ring calculation work.

\section{Conclusion}
\label{sec:conclusion}

Total persistent $H_1$ of bond percolation on $\{3,q\}$ disks, $q\ge7$, is given to first order in $1/\lambda(q)$ by \eqref{eq:h1-derived}, with no fitted parameters and a closed-form large-curvature intercept $5/4-2\pi/(3\sqrt3)$. The derivation rests on an exact Euler-characteristic identity, the boundary fraction $1-1/\lambda$, and the substitution \eqref{eq:substitution} describing how each ring is built from the last.

What remains open is the $O(\lambda^{-2})$ term, measured at about $-0.01/\lambda^2$ (Table~\ref{tab:mu}). It collects bridging through two rings, correlations between bridges at neighboring rim breaks, and the dependence on the arrangement of vertex types, all local computations of the same kind as \eqref{eq:beta}; there is no obstacle in principle to carrying the expansion one order further. Whether the full series resums to something closed is a separate question. The method applies to any $\{p,q\}$ family built by ring growth, with $\lambda$, the substitution, and the seam structure replaced by their analogues.

Finally, the quantity computed here is a property of the abstract complex. The motivating question, how much a tree-based embedding of percolated hierarchical data is distorted by loops, concerns the embedded point cloud and whatever model is trained on it. Embedding clusters by a branching random walk along a spanning tree, measuring the resulting holonomy at cycle-closing edges, and asking whether its dependence on curvature carries into the features a sparse autoencoder learns \citep{bricken2023,cunningham2023}, is a natural next step. A preliminary pipeline exists but its results are not yet stable enough to report.

\section*{Code and data availability}

All code and data are available at \url{https://github.com/ztreisman/curvature-percolation-homology}. The identity \eqref{eq:exact-identity} was checked against GUDHI's computed total persistence to $10^{-12}$ on a $\{3,7\}$ disk with $V=617$; scripts verifying the substitution \eqref{eq:substitution}, the bridging probability \eqref{eq:beta}, and each entry of Tables~\ref{tab:mu} and~\ref{tab:h1} are in the \texttt{experiments/} directory.

\bibliographystyle{plainnat-hyphen}
\bibliography{references}

\end{document}